\documentclass[aps,pre,floatfix,nofootinbib,notitlepage,10pt]{revtex4-1}

\usepackage{graphicx}
\usepackage{amsmath,amssymb,amsfonts}
\usepackage{bbm}
\usepackage{color}
\usepackage{leftidx}

\newcommand{\Br}{{\bf r}}

\newcommand{\Bu}{{\bf u}}
\newcommand{\Bv}{{\bf v}}

\newcommand{\BF}{{\bf F}}

\newcommand{\Bn}{{\bf n}}

\newcommand{\Bdelta}{\boldsymbol{\delta}}

\newcommand{\BG}{{\bf G}}

\newcommand{\tenG}{{\bf G}}

\begin{document}

\title{Permeability of immobile rings of membrane inclusions to in-plane flow}

\author{Yulia Sokolov}

\author{Haim Diamant}
\affiliation{Raymond and Beverly Sackler School of Chemistry, Tel Aviv
  University, Tel Aviv 6997801, Israel}

\date{11 March 2019}

\begin{abstract}
We study the flow of membranal fluid through a ring of immobile
particles mimicking, for example, a fence around a membrane corral. We
obtain a simple closed-form expression for the permeability
coefficient of the ring as a function of the particles' line
fraction. The analytical results agree with those of numerical
calculations and are found to be robust against changes in particle
number and corral shape. From the permeability results we infer the
collective diffusion coefficient of lipids through the ring and
discuss possible implications for collective lipid transport in a
crowded membrane.
\end{abstract}

%\pacs{82.70.Dd, 87.80.Cc}
% 82.70.Dd: Colloids
% 87.80.Cc: Optical trapping

\maketitle

\section{Introduction}
\label{sec_intro}

Numerous studies continue to unravel the complex structure and
dynamics of biological membranes
\cite{Lyman2018,Shi2018,Chein2019,Krapf2018,Guigas2016}. The basic
component of biomembranes, the fluid lipid bilayer, contains a variety
of embedded objects such as mobile and immobile proteins and domains
of different lipid composition. The resulting highly crowded
environment affects the dynamics of all its constituents.

Indeed, it has been known for over three decades that the cortical cytoskeleton hinders the diffusion of membrane proteins and lipids \cite{Sheetz1980,Golan1980,Tsuji1986,Tsuji1988}. Later studies connected this inhibition with the subdivision of the membrane into domains (``corrals''), enclosed by immobile proteins (``fences and pickets'') \cite{Sako1994,Sako1995,Salome1998,Ritchie2003,Kusumi2006,Lee2018,Krapf2018}. In such a compartmentalized membrane, a mobile protein or lipid diffuses within one of the domains until it escapes and moves into the next domain, and so on. Many experimental and theoretical studies have been devoted to the resulting anomalous dynamics \cite{Shutz1997,Weigel2011,Skaug2011,Krapf2015,Golan2017,Saxton1995,Saxton1997,Leitner2000,Brown2000,Niehaus2008,Auth2009,Bresslof2009,Holcman2011,Jeon2016,Metzler2016,Javanainen2017,Sumi2017}.

The most common techniques used in the experimental studies are fluorescence recovery after photobleaching (FRAP) \cite{Sheetz1980,Golan1980,Tsuji1986,Salome1998,Lee2018}, single-particle tracking (SPT) \cite{Saxton1997,Shutz1997,Ritchie2003,Weigel2011,Skaug2011,Golan2017,Lee2018}, and fluorescence correlation spectroscopy (FCS) \cite{Wawrezinieck2005,Destainville2008,Machan2010}. Note that these measurements follow either single molecules (SPT and FCS) or dilute dispersions of fluorescent probes (FRAP), and hence do not yield the collective dynamics of the membranal fluid. If we want to write the diffusion equation for the dense bulk fluid of lipids, the relevant diffusion coefficient is the collective one, which can differ by orders of magnitude from the single-lipid coefficient \cite{Vattulainen2005,Falck2008}. It is of interest, therefore, to account for collective lipid transport, which is exepcted to be strongly affected by membrane compartmentalization.

In the present work we attempt to account for the collective transport of lipids into or out of a membrane corral. We address the physics of the transport hindrance due to the corral boundary, going beyond the schematic descriptions offered by previous works, such as a potential barrier or a stochastic gate \cite{Saxton1995,Leitner2000,Brown2000,Niehaus2008,Auth2009,Bresslof2009,Holcman2011,Jeon2016,Metzler2016,Javanainen2017,Sumi2017}. The collective diffusion coefficient across the fence can be inferred from the fence's permeability coefficient, since the two are directly related via the fluctuation-dissipation theorem.

In order to calculate the permeability coefficient we start, as in most theories of membrane flow, from the classical work of Saffman and Delbr\"uck (SD) \cite{SD,Saffman}. In this simplified model the membrane is perceived as a flat slab of homogeneous viscous fluid embedded in a much less viscous solvent, and the proteins ---  as cylindrical inclusions spanning the thickness of the membrane.\footnote{\setlength{\baselineskip}{12pt}Due to the relatively low viscosity of the solvent, the possible protrusion of the inclusion out of the membrane usually has an insignificant effect on the dynamics.} Over the years the SD model has been applied to diverse problems\cite{Brown}, among which are the diffusion of single proteins \cite{SD,Hughes}, the hydrodynamic coupling of two or several proteins \cite{Bussel1992,Bussel1994,Naomi,Noruzifar}, the influence of inclusions on the effective viscosity of the membrane \cite{Bussel,Naomi,Henle}, and the effect of the inclusion's shape on its diffusion \cite{LevineMob,Ramachandran,CamleyBrown}. In particular, the presence of a solid substrate \cite{Stone, OppenDiam2010,Seki,Komura} or immobile obstacles \cite{Bussel,OppenDiam2011} was shown to have a strong effect on membrane dynamics. 

A more realistic theoretical description of membrane dynamics would require an understanding of lipid transport through membrane obstructions. This would be also essential for predicting and interpreting experimental results. As a step in this direction, we adapt the SD model to treat membranal flow through a ring-like fence of immobile inclusions.

In the next section we describe our model in detail. In section \ref{sec_res} we present the results of numerical calculations, analytical derivation of the permeability coefficient, and the resulting collective diffusion of lipids. Finally, in section \ref{sec_disc} we discuss the results and their implications, and indicate the limitations of our model.

\section{Model}
\label{sec_model}

\begin{figure}[tbh]
%\vspace{0.2cm}
 \centering
 \includegraphics[width=0.5\columnwidth]{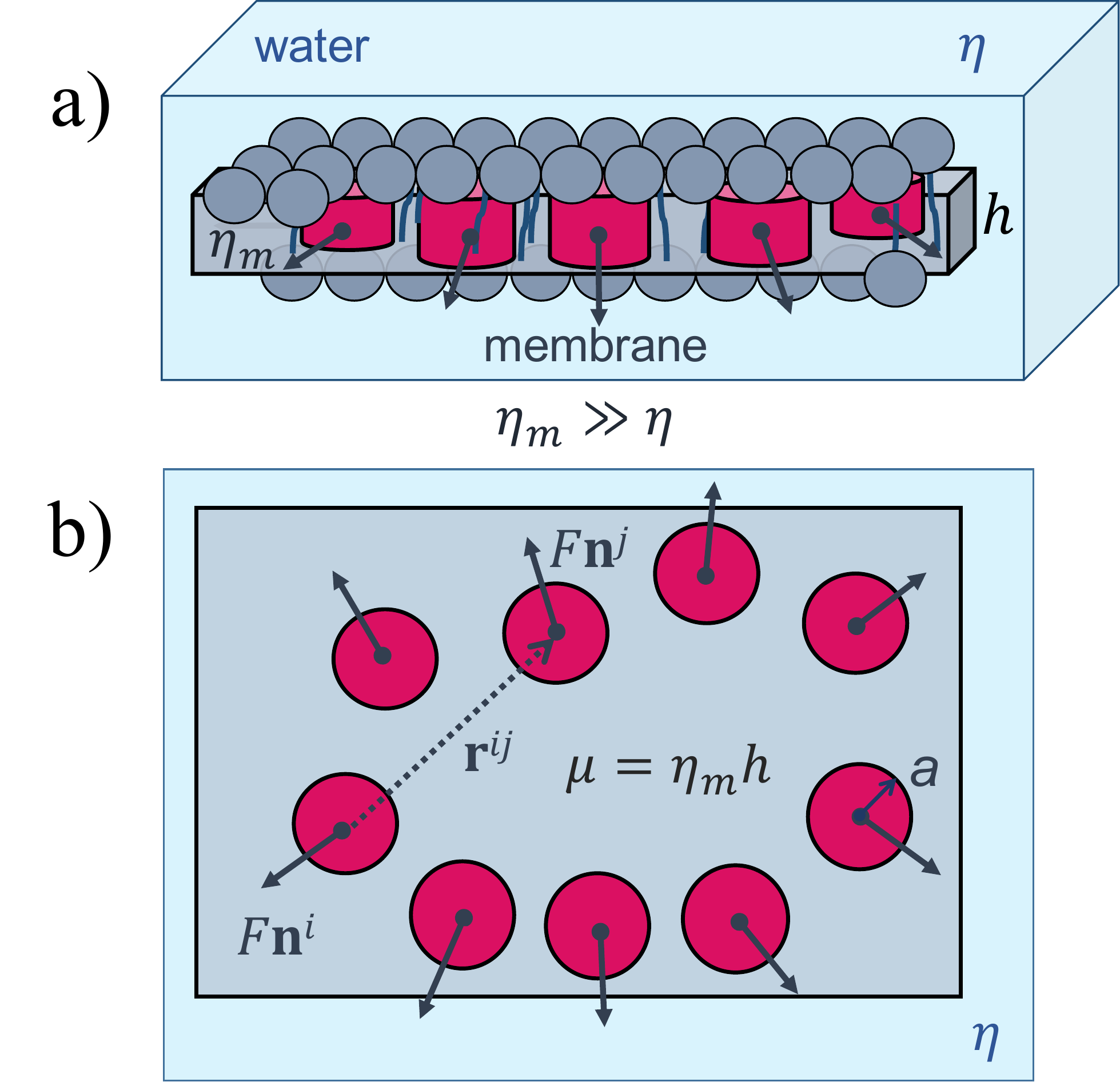}
  \caption{Schematic illustration of a corral in a membrane immersed in water. Panels a) and b) present the side and top views, respectively.}
\label{fig:memring}
\end{figure}

We consider an assembly of $N$ membrane proteins arranged equidistantly in a ring-like structure of arbitrary shape and perimeter $L$  (see Fig.~\ref{fig:memring}). The proteins are represented as cylindrical inclusions of radius $a$, embedded in a membrane of thickness $h$ and viscosity $\eta_m$, within a surrounding aqueous solution of viscosity $\eta \ll \eta_m$. We define an effective 2D viscosity, $\mu=\eta_m h$, the ring density (line fraction) $\phi=2 Na/L$, and the cutoff length (the SD length) $\kappa^{-1}=\mu/(2\eta)$ \cite{Saffman, SD}$^{,}$\footnote{\setlength{\baselineskip}{12pt}If a membrane is surrounded by two different fluids with different viscosities $\eta_1$ and $\eta_2$, then $\kappa^{-1}=\mu/(\eta_1+\eta_2)$.}. In reality, because of the large viscosity of the membrane, the SD length is about $2$--$3$ orders of magnitude larger than the membrane thickness, i.e., of micron scale \cite{Brown}, whereas the corral size is of order $0.1$ $\mu$m \cite{Salome1998,Kusumi2006}. For simplicity we assume $L \ll \kappa^{-1}$. As we shall see, in this limit, the results are insensitive to the value of $\kappa^{-1}$.

In the system under consideration, the ring is immobile. A pressure difference $\Delta p$ gives rise to a lipid flow $\Bu$, which is perturbed in the vicinity of each immobile inclusion. Since the proteins interact hydrodynamically, this perturbation around each individual protein $i$ is affected by all other proteins $j \neq i$. In order to keep each protein $i$ in place, a restoring force $\BF^i$ is applied, for example, by the cytoskeleton. In the continuum limit, the permeability $P$ of the corral to lipid flow is defined as the ratio between the normal flow and the pressure difference, $P = \langle \Bu \cdot \Bn \rangle /\Delta p$ (Darcy's law), where $\Bn$ is a unit normal to the corral boundary at a given point, and the average is taken over the perimeter. For a discrete chain of particles we define the normal to the polygonal ring at a given particle, $\Bn^i$, as the unit vector along the mean direction of the normals to the adjacent two chords.

For the sake of analytical convenience, we reconstruct the problem in the following way. Instead of immobile particles within an imposed flow, we may consider the near-equivalent problem of mobile particles driven by normal forces $\BF^i=F \Bn^i$ in an ambient membrane (Fig.~\ref{fig:memring}). This reciprocal equivalence is exact in the limit of an infinite, straight-line arrangement, due to translational invariance (momentum conservation). As we shall see, our results for the permeability (both analytical and numerical) are insensitive to the curvature of the ring, which proves the validity of this approximation. In the inverted description, the flow caused by the motion of each driven particle influences the motions of all other particles in the assembly. The permeability then can be rewritten as the ratio between the average normal velocity of the particles and the total normal force per unit length,
\begin{equation}
P=\frac{L}{NF} \langle \Bv^i \cdot \Bn^i \rangle=\frac{2a}{\phi F} \langle \Bv^i \cdot \Bn^i \rangle.
\label{permdef}
\end{equation}
In order to find the average over the $N$ particles, the normal velocity of each particle is to be calculated as a function of the positions of all particles.

Our primary goal is to calculate the permeability $P$ as a function of line fraction $\phi$ and particle number $N$. Note that we work within the continuum limit for the membrane, under the rough assumption that the size of the lipids is negligible relative to the gaps between the corral proteins. 

\section{Results}
\label{sec_res}

\subsection{Numerical calculation}
\label{sec:numres}

The velocity of particle $i$, given the positions of all other particles in the ring, is given by
\begin{eqnarray}
&&\Bv_{\alpha}^i=F\left(B_s \Bn_{\alpha}^i+ \sum_{j \neq i} \tenG_{\alpha \beta}^{ij} \Bn_{\beta}^j \right), \label{vel}\\
&&B_s=\frac{1}{4 \pi \mu}\left(\ln \frac{2}{\kappa a} -\gamma\right).
\label{self}
\end{eqnarray}
Equation (\ref{vel}) contains two terms. The first is the velocity of particle $i$ due to the force acting on it, where $B_s$ is the self-mobility as calculated by SD \cite{SD}, and $\gamma \simeq 0.58$ is Euler's constant. The second term represents the hydrodynamic interactions between particles $i$ and $j \neq i$, positioned at $\Br^i$ and $\Br^j$, where $\tenG_{\alpha \beta}^{ij} \equiv \tenG_{\alpha \beta} (\Br^j-\Br^i)$ is the coupling mobility tensor and $\alpha, \beta=x,y$ are membrane coordinates.

\begin{figure}[tbh]
%\vspace{0.2cm}
 \centering
 \includegraphics[width=0.6\columnwidth]{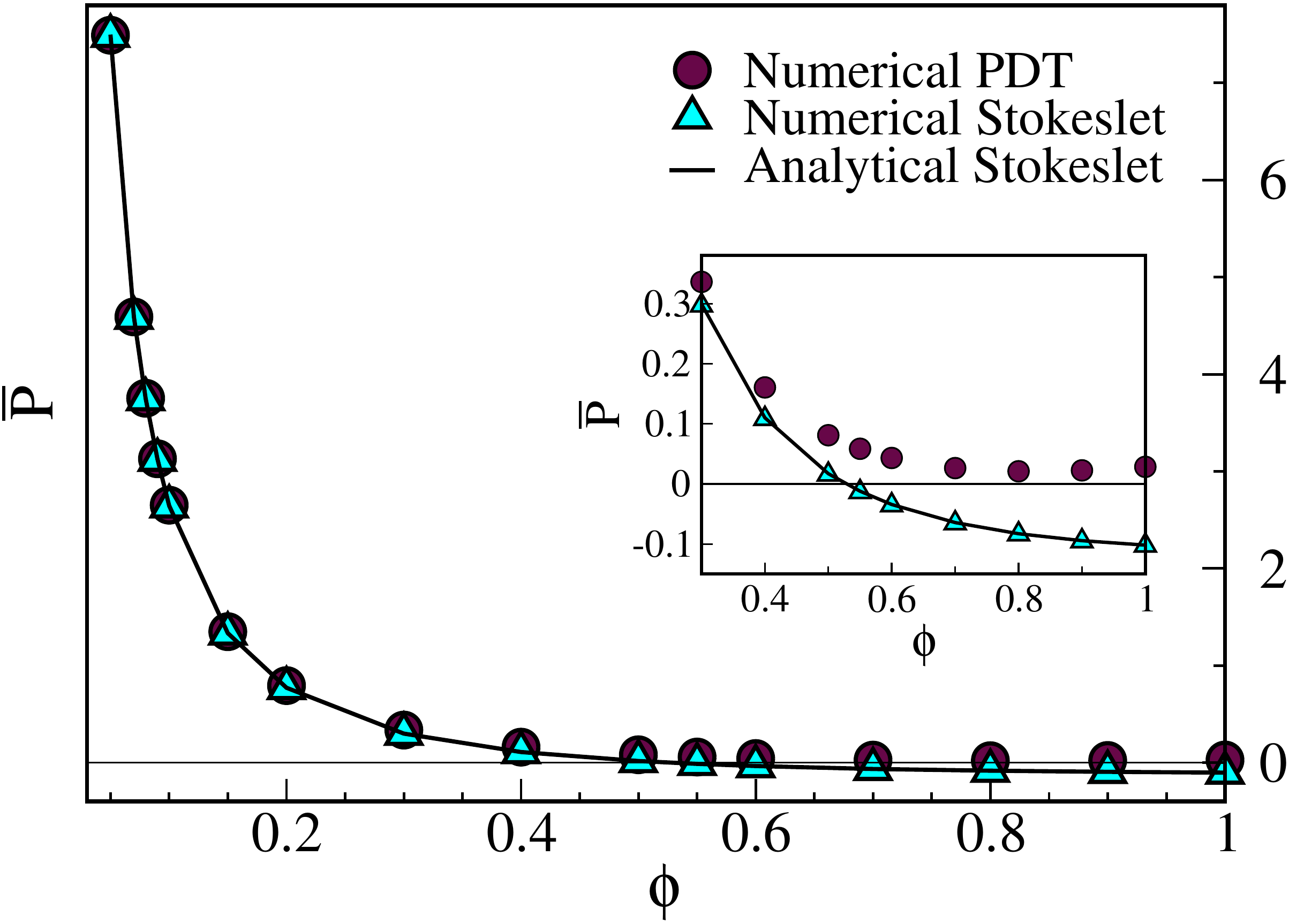}
\caption{Permeability coefficient as a function of line fraction for a circular corral of $N=100$ particles. The coefficient is normalized according to $\bar{P}=(\mu/a)P$.  The line fraction $\phi$ is varied while keeping the particle number constant by changing the protein size according to $a=\phi L/(2 N)$. Triangles show results of numerical summation using the interaction tensor $\BG$ [Eq.~(\ref{Sten})] (Stokeslet limit), and circles --- using the corrected, positive-definite tensor (PDT), $\BG^+$ [Eq.~(\ref{YHten})]. In the numerics we normalize the parameters such that $L=2 \pi$ and $\kappa=10^{-3}$. The solid line presents the analytical result, which is independent of $L$ and $\kappa$ [Eq.~(\ref{P})]. The inset focuses on the failure of the tensor $\BG$ for $\phi \gtrsim 0.5$.}
\label{fig:simpdth}
\end{figure}

We consider two possible approximations for the interaction tensor $\tenG$. The simple approximation assumes the particles to be point-like objects (the Stokeslet limit) \cite{Bussel1992,Levine2002,Naomi},
\begin{eqnarray}
\BG_{\alpha\beta}(\Br) = \frac{1}{4 \pi \mu}
  \left[ \left(\ln \frac{2}{\kappa r} -\gamma -\frac{1}{2} \right)
    \Bdelta_{\alpha \beta} + \frac{\Br_{\alpha} \Br_{\beta}}{r^2}
    \right],
  \ \ \ a\ll r \ll \kappa^{-1},
\label{Sten}
\end{eqnarray}
where $r=|\Br^j-\Br^i|$. As we have shown in a recent publication
\cite{SD2018}, this tensor fails above a relatively large density,
$\phi > \phi_c \simeq 0.5$, producing inward motion under outward
force. When this tensor is used for a circular ring of equidistant
particles, we get negative permeabilities above that density, as shown
in the inset of Fig.~\ref{fig:simpdth} (light triangles) and as
calculated below. In Ref.~\citenum{SD2018} we derived an improved
interaction tensor, overcoming the problem of negative response,
\begin{equation}
\BG^+_{\alpha \beta}(\Br)  = \frac{1}{4 \pi \mu}
  \left [\left(\ln \frac{2}{\kappa r} -\gamma -\frac{1}{2} +\frac{a^2}
    {r^2}\right)
    \Bdelta_{\alpha \beta} + \left(1-\frac{2 a^2}{r^2}\right)
    \frac{\Br_{\alpha} \Br_{\beta}}{r^2}\right],
  \ \ a \ll r \ll \kappa^{-1}.
\label{YHten}
\end{equation}
When this corrected tensor is substituted in
Eqs.~(\ref{permdef}) and (\ref{vel}), we find strictly positive
permeability coefficients (Fig.~\ref{fig:simpdth}, dark circles) for
the same circular ring of particles. These results are physical, but
not exact. For instance, for a sealed corral, at $\phi =1$, we get a
small positive permeability, while in reality the permeability should
vanish.

\begin{figure}[tbh]
%\vspace{0.2cm}
 \centering
 \includegraphics[width=0.5\columnwidth]{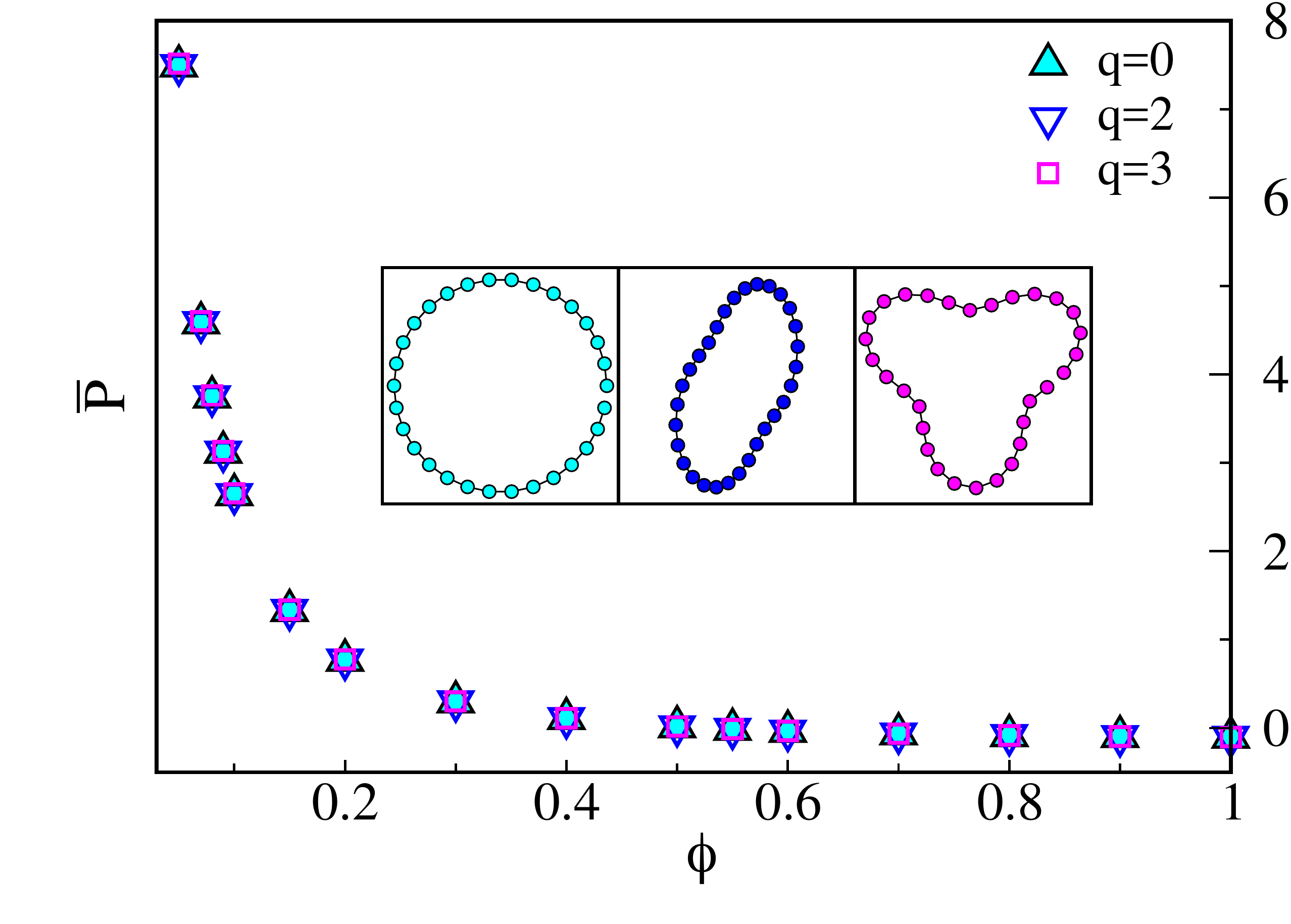}
  \caption{Normalized permeability coefficient $\bar{P}=(\mu/a)P$ as a function of line fraction $\phi$ for particle number $N=30$ and three different ring shapes, obtained by numerical summation in the Stokeslet limit. The three shapes correspond to three modulation modes, according to Eq.~(\ref{Rshape}) with $\rho=0.3$. The shapes are constructed such that their perimeters are the same, $L=2 \pi$.}
\label{fig:shape}
\end{figure}

In order to check the sensitivity of the results to the shape of the corral, we study two additional ring shapes, using the function
\begin{equation}
\tilde{R}(\theta)=R \left[1+\rho \sin (q \theta)\right],
\label{Rshape}
\end{equation}
and examining two modulations, $q=2,3$, in addition to the basic
circular case, $q=0$. See the shapes in Fig.~\ref{fig:shape}. The same
number of particles should correspond to the same line fraction in
order to achieve a meaningful comparison between the three
shapes. Therefore, for each shape, we calculate the value of $R$ such
that the perimeter $L$ remains unchanged. As shown in
Fig.~\ref{fig:shape}, the permeabilities of the three rings are
indistinguishable. The reason will become clear shortly. This is the
conclusion when using either $\BG$ or $\BG^+$.

\subsection{Analytical calculation}
\label{sec:anres}

\begin{figure}[tbh]
%\vspace{0.2cm}
 \centering
 \includegraphics[width=0.5\columnwidth]{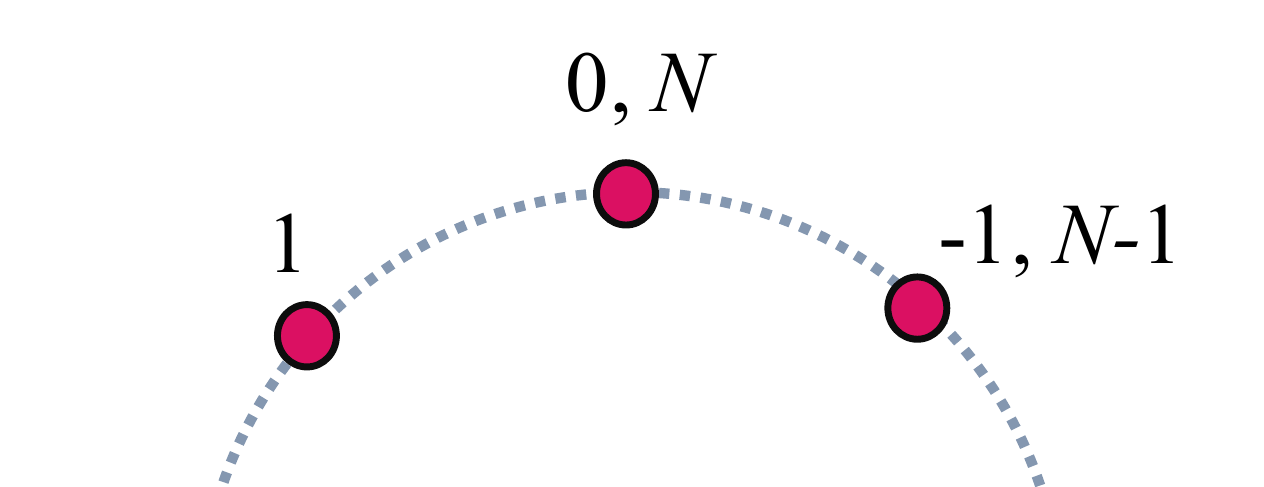}
  \caption{Numbering the particles in the ring.}
\label{fig:numring}
\end{figure}

We calculate analytically the expression for the permeability in the Stokeslet limit and compare it to the numerical results. This calculation is strictly valid in the continuum limit, $N \rightarrow \infty$ and $a \rightarrow 0$ such that $Na=L\phi/2$ is finite, but we shall see that practically it works for remarkably small particle numbers. We number the particles along the ring by the index $m \in[0,N-1]$ and take particle $m=0$ to be identical to $m=N$ to close the ring. Thus, particle $m=N-1$ can be marked also as $m=-1$, etc. See Fig.~\ref{fig:numring}. According to Eq.~(\ref{vel}) and focusing on particle $m=0$, we need to calculate the following sum:
\begin{equation}
S_{\alpha}\equiv\sum^{N-1}_{m=1}\BG^{m,0}_{\alpha\beta} \Bn^m_{\beta},
\label{sum}
\end{equation}
where $\BG^{m,0}_{\alpha \beta}\equiv\BG_{\alpha\beta}(\Br^m-\Br^0)$. This sum accounts for the hydrodynamic interactions of a given particle ($m=0$) with the rest of the particles in the ring ($m \neq 0$). 

In the first step we transform the discrete sum into a continuous
integral using the Euler-Maclaurin formula:
\begin{eqnarray}
&&S_{\alpha}\simeq \int^{N-1}_{1}dm\, \BG^{m,0}_{\alpha\beta}\,
  \Bn^m_{\beta}+{\cal R}^{(1)}+{\cal R}^{(2)},\\ &&{\cal
    R}^{(1)}=\frac{1}{2}\left[\BG^{1,0}_{\alpha\beta}\,\Bn^1_{\beta}+\BG^{-1,0}_{\alpha\beta}\,\Bn^{-1}_{\beta}\right],\nonumber
  \\ &&{\cal
    R}^{(2)}=-\frac{1}{12}\left[\partial_m\BG^{1,0}_{\alpha\beta}\left.|_{m=1}\,\Bn^1_{\beta}
    -\partial_m\BG^{-1,0}_{\alpha\beta}\right.|_{m=-1}\,\Bn^{-1}_{\beta} \right].\nonumber
\label{sumtoint}
\end{eqnarray}
Including the two leading corrections ${\cal R}^{(1)}$ and ${\cal
  R}^{(2)}$ is sufficient to fit the numerical results.

In the second step we exploit the fact that $\BG$ is divergenceless,
$\partial_{\alpha}\BG_{\alpha\beta}=0$, due to the assumed
incompressibility of the lipid flow. Since we apply normal forces
along the entire closed boundary of the domain, according to the
divergence theorem, the full integral along the ring vanishes,
$\int^N_0dm\BG^{m,0}_{\alpha\beta} \Bn^m_{\beta}=0$. Hence,
\begin{equation}
\int^{N-1}_{1}dm\BG^{m,0}_{\alpha\beta}\,\Bn^m_{\beta}=-\int^{1}_{-1}dm\BG^{m,0}_{\alpha\beta}\,\Bn^m_{\beta}.
\label{divint}
\end{equation}
In this way we have reduced the summation of many long-range interactions between a given particle and the rest of the ring to an integration over a much smaller domain around the particle, containing only its two nearest neighbors.

In the third step we obtain the structural functions $\Br(m)$ and $\Bn(m)$. For $N \gg 1$ we can assume a small curvature, %$c<<1,$
in which case the neighbors of the given particle are positioned at a distance $\ell_0\simeq L/N$, and
\begin{equation}
  r^{m,0}=|\Br^m-\Br^0|\simeq |m|\ell_0,
  \ \ \Bn^m \simeq \Bn^0.
 \label{dist}
 \end{equation}
 Note that now $m$ is restricted to the interval $[-1,1]$.
%In the same limit we expand for $\Bn^m=\Bn(m)$,
%\begin{eqnarray}
%\Bn(m)&=&\Bn(0)+m\Bn'(0)+\frac{m^2}{2}\Bn''(0)=\nonumber \\
%&=&\Bn(0)+mc(0)\Bt(0)+\frac{m^2}{2}c'(0)\Bt(0)-\frac{m^2}{2}c^2(0)\Bn(0),
%\label{norm}
%\end{eqnarray}
%where prime denotes a derivative with respect to $\Bn$, $\partial_m$. In this equation we have defined the unit tangent to the ring at a given position $m$, $\Bt(m)$, along the continuous interval %$m\in[-1,1]$, and a dimensionless curvature $c(m)$, such that $\Bn'=c\Bt$ and $\Bt'=-c\Bn$ (See Fig.~\ref{fig:scheme}).

In the fourth and last step we substitute Eq.~(\ref{dist}) % and (\ref{norm})
into Eq.~(\ref{sumtoint}). The resulting normal component of the sum can be written as
\begin{eqnarray}
&&S_n \equiv S_{\alpha}n_{\alpha}=-\int^1_{-1}dm\,g(|m|\ell_0)+g(\ell_0)-\frac{\ell_0}{6}g'(\ell_0),
\label{finS}
\end{eqnarray}
%\begin{eqnarray}
%&&S_n=-B_s\left[{\cal I}_n(\ell_0)-\frac{c_0^2}{2}{\cal J}_n(\ell_0)\right], \\
%&&{\cal I}_n(\ell_0)=\int^1_{-1}dm\,g(|m|\ell_0)-g(\ell_0), \nonumber \\
%&&{\cal J}_n(\ell_0)=\int^1_{-1}dm\,m^2\,g(|m|\ell_0)-g(\ell_0), \nonumber
%\label{finS}
%\end{eqnarray}
where $g(r)$ is the isotropic part of the interaction tensor $\BG$,
\[
g(r)=\frac{1}{4 \pi \mu}\left(\ln\frac{2}{\kappa r}-\gamma-\frac{1}{2}\right).
\]

Let us consider for a moment a circular ring such that the velocities of all particles are equal by symmetry. Performing the integration in Eq.~(\ref{finS}) while using Eqs.~(\ref{permdef}),~(\ref{vel}) and (\ref{sum}), we obtain our central result for the permeability of the corral,
\begin{equation}
P(\phi)=\frac{a}{2 \pi \mu} \frac{1}{\phi}\left(\ln \frac{2}{\phi}-\frac{4}{3}\right).
\label{P}
\end{equation}
Note that the dependence on the cutoff length $\kappa^{-1}$ has disappeared.

\begin{figure}[tbh]
%\vspace{0.2cm}
 \centering
 \includegraphics[width=0.5\columnwidth]{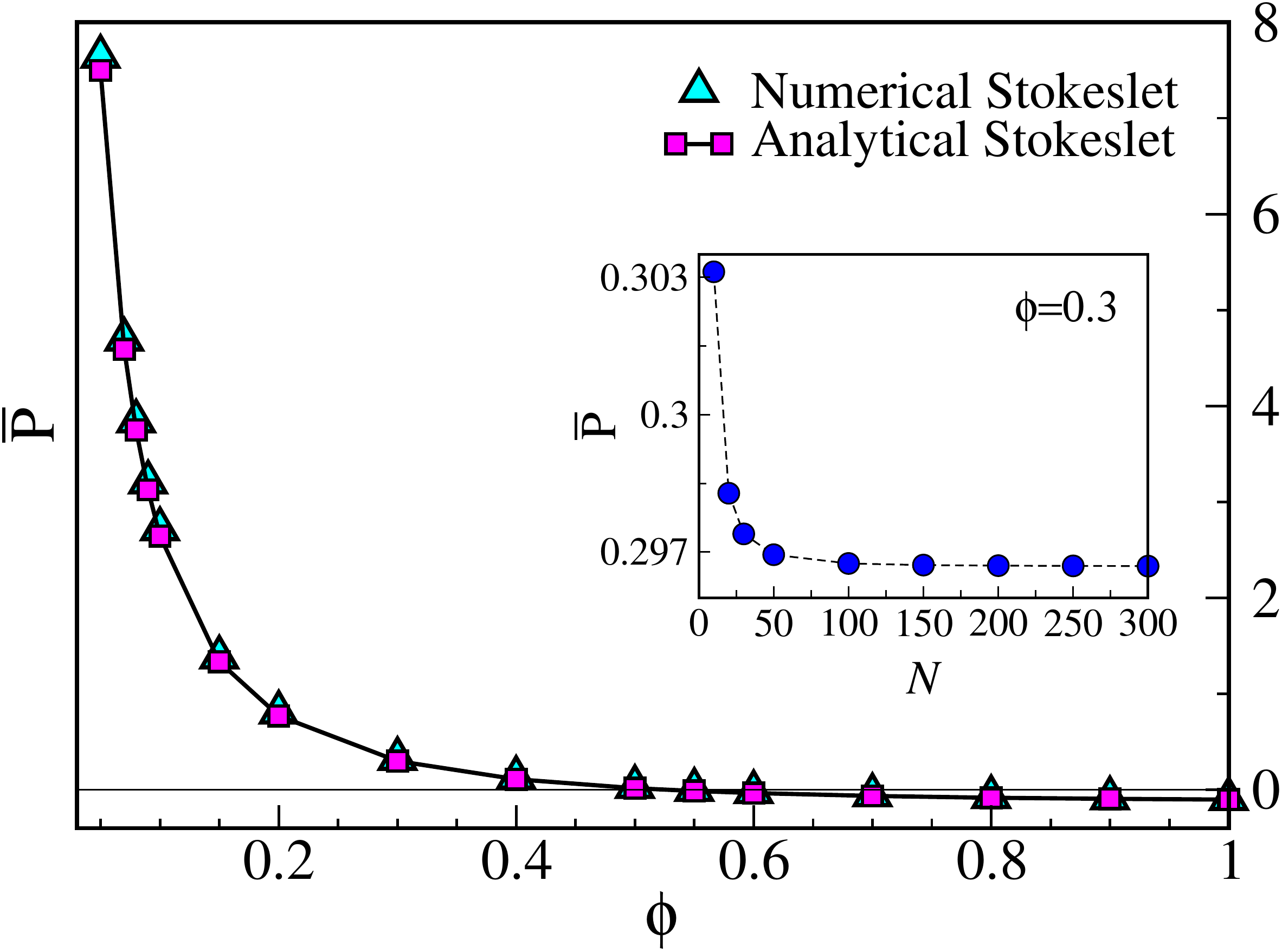}
  \caption{Normalized permeability coefficient $\bar{P}=(\mu/a)P$ as a function of line fraction $\phi$ for small particle number ($N=5$). Triangles show the numerical result in the Stokeslet limit and the squares with solid line show the analytical result. Note that the relative errors remain small even for $N=5$. The inset presents the dependence of the numerical results on particle number for $\phi=0.3$ and shows the fast convergence of $\bar{P}$ to its continuum value as the particle number increases.}
\label{fig:permN}
\end{figure}

Despite the aforementioned assumptions, this result is robust against changes in particle number and ring shape. The solid line in Fig.~\ref{fig:simpdth} shows excellent agreement with the Stokeslet numerical results for $N=100$. Fits of similar quality are obtained for smaller particle numbers, as demonstrated in Fig.~\ref{fig:permN}.

Concerning the non-circular rings, studied in the preceding section, the analytical calculation clarifies why the numerical results presented in Fig.~\ref{fig:shape} are insensitive to the ring shape. The divergence theorem is valid for any closed shape and, as we have just shown, reduces the interactions to the small immediate vicinity of an arbitrarily chosen particle, thus making the rest of the ring immaterial.

Recall that results obtained within the Stokeslet approximation become unphysical above a certain density $\phi_c$ \cite{SD2018}. This unphysical behavior is overcome by the corrected interaction tensor $\BG^+$ of Eq.~(\ref{YHten}). Unfortunately, the procedure leading to Eq.~(\ref{P}) cannot be applied to $\BG^+$, since the extra $1/r^2$ term causes divergence of the integral in Eq.~(\ref{finS}). The inset of Fig.~\ref{fig:simpdth} compares the numerical and analytical results and shows that for $\phi \gtrsim 0.4$ numerical summation with the $\BG^+$ tensor should be used. In addition, Eq.~(\ref{P}) allows us to calculate $\phi_c$ analytically. We find that $P(\phi)>0$ for $\phi<\phi_c=2 e^{-4/3}\simeq 0.527$.

In the opposite limit, $\phi \rightarrow 0$, Eq.~(\ref{P}) gives indefinitely large permeability. In reality, there is a finite $\phi$ below which Darcy's law no longer holds, and our theory becomes irrelevant.

\subsection{Consequences for collective lipid diffusion}
\label{sec_diff}

One of the important biological implications of the permeability is for the collective (gradient) diffusion of lipids across a fence of membrane inclusions, under the continuum assumption that the distance between the obstacles is much larger than the lipid size. While the permeability coefficient is the ratio between flow velocity and pressure difference across the fence, the gradient diffusion coefficient $D$ is the ratio between flux and concentration gradient. The two are related via
\begin{equation}
D(\phi)=2a K P(\phi),
\label{D}
\end{equation}
where $K=c(\partial p/\partial c)_T$ is the isothermal compression modulus of the membrane, $c$ is the lipid concentration, and $2 a$ is the assumed thickness of the fence\footnote{\setlength{\baselineskip}{12pt}Equation (\ref{D}) generalizes the known linear relation between permeability and diffusion coefficient in gases \cite{Crank1975}.}. Thus, Eqs. (\ref{P}) and (\ref{D}) give a simple prediction for the collective diffusion coefficient across a ring-like or an infinitely long, linear fence as a function of the fence's density.

Substituting realistic values, $a=3$ nm, $K=200$ erg/cm$^2$,\cite{Marsh} $\phi=0.3$, and $\mu=10^{-6}$ erg$\cdot$s/cm$^2$ (surface poise) \cite{Brown}, we get $D \sim 10^3$ $\mu$m$^2$/s. This value is two orders of magnitude larger than known self-diffusion coefficients of a single lipid \cite{Kusumi2006}, as expected for the collective diffusion coefficient in a dense fluid \cite{Vattulainen2005,Falck2008}. The high density of the lipid fluid enters through the relatively large value of $K$. According to liquid theory, $D \sim D_s/S(k \rightarrow 0)$, where $D_s$ is the self-diffusion coefficient of a single lipid, and $S(k)$ is the structure factor as a function of wavevector $k$. In the limit of $k \rightarrow 0$,  $S \sim 1/K$, \cite{Vattulainen2005,Liquid} and, therefore, $D \sim K$, in line with Eq.~(\ref{D}).

As mentioned in the previous section, at sufficiently low line
fraction the expression (\ref{P}) becomes invalid. In this limit $D$
should saturate to its value in a neat membrane.

\section{Discussion}
\label{sec_disc}

In this work we have derived the permeability coefficient of a ring of immobile particles, embedded in a fluid membrane. Earlier studies highlighted the distinctive hydrodynamic behavior of the membrane as a quasi-two-dimensional fluid \cite{Brown,OppenDiam2011}, expressed, for example, in the appearance of logarithmic functions \cite{Saffman}. In the present work this anomaly is reflected by the logarithmic dependence on density in Eq.~(\ref{P}).

Our results have a broad range of validity with respect to the density of the ring, its shape, and the number of constituent particles. This, in particular, justifies our initial assumption concerning the equivalence between immobile and forced particles regardless of curvature. The resulting permeability is a purely local, intensive property, which is not self-evident in the present case in view of the very long-range (logarithmic) hydrodynamic interactions. In addition, the insensitivity to shape implies that the results also apply for long lines of immobile inclusions away from their edges.

The expression for the permeability allows to calculate the collective
diffusion coefficient of lipids into and out of a corral, according to
Eq.~(\ref{D}). In a membrane crowded with such corrals, diffusion
through their boundaries, consisting of immobile inclusions, should
become the rate-limiting process for lipid transport. Since the
permeability undergoes strong suppression with increasing line
fraction of proteins in the corral fences (see
Fig.~\ref{fig:simpdth}), we expect a similar suppression of the
collective diffusion coefficient of lipids over length scales larger
than the corral size, compared to a neat membrane
\cite{Shi2018,Chein2019}. Note that this suppression does not require
a high area fraction of immobile inclusions, but just a high line
fraction at the boundaries of the corrals.

The simplified model presented here has several limitations when applied to biomembranes: (a) the ring is never isolated in reality, but rather surrounded by other mobile and immobile intramembranal objects; (b) the membrane is not a continuous fluid, since the lipids have a finite size which is not necessarily negligible compared to the corral proteins; (c) the corral proteins need not be of the same lateral size or arranged equidistantly; (d) the inclusions are never perfectly immobile; and (e) the condition $L \ll \kappa^{-1}$ is not strictly fulfilled in reality. Nonetheless, we believe that our theory should capture the essential characteristics of lipid transport through a ring-like assembly of obstacles.

\begin{acknowledgments}
We thank Tomer Goldfriend for a helpful discussion. This study was supported by the Israel Science Foundation (Grant  Nos.\ 164/14 and 986/18).
\end{acknowledgments}

\end{document}